\documentclass{emulateapj}
\usepackage{apjfonts}

\newcommand{\be}{\begin{equation}}
\newcommand{\bel}[1]{\begin{equation}\label{eq:#1}}
\newcommand{\ee}{\end{equation}}
\newcommand{\bd}{\begin{displaymath}} 
\newcommand{\ed}{\end{displaymath}}   
\newcommand{\bea}{\begin{eqnarray}}
\newcommand{\beal}[1]{\begin{eqnarray}\label{eq:#1}}
\newcommand{\eea}{\end{eqnarray}}
\newcommand{\eqlab}[1]{\label{eq:#1}}
\newcommand{\eqref}[1]{\ref{eq:#1}}

\newcommand{\Rs}{{R_{\rm S}}}
\newcommand{\pder}[2]{{\partial #1 \over \partial #2}}
\newcommand{\Teff}{{T_{\rm eff}}}
\newcommand{\zth}{^{(0)}}
\newcommand{\fst}{^{(1)}}
\newcommand{\Bdot}{{\dot B}}

\newcommand{\kappaT}{{\kappa_{\rm T}}}

\newcommand{\rad}{_{\rm rad}}
\newcommand{\con}{_{\rm con}}

\newcommand{\tot}{_{\rm tot}}
\newcommand{\Spitzer}{\lambda_{\rm c}}
\newcommand{\Ne}{ n_{\rm e}}
\newcommand{\Np}{ n_{\rm p}}
\newcommand{\grad}{ g_{\rm rad}}
\newcommand{\Mns}{{M_{\rm ns}}}
\newcommand{\Msol}{{M_{\odot}}}
\newcommand{\Rns}{{R_{\rm ns}}}

\newcommand{\Fobs}{{F_{\rm obs}}}
\newcommand{\nuobs}{{\nu_{\rm obs}}}
\newcommand{\Rinf}{{R_{\infty}}}
\newcommand{\Tinf}{{T_{\infty}}}
\newcommand{\RS}{{R_{\rm S}}}
\newcommand{\gs}{{g_{\rm s}}}

\def\Chandra{${\it Chandra}$\ }

\newcommand{\Msun}{\ifmmode {M_{\odot}}\else${M_{\odot}}$\fi}
\newcommand{\Lsun}{\ifmmode {L_{\odot}}\else${L_{\odot}}$\fi}
\newcommand{\Rsun}{\ifmmode {R_{\odot}}\else${R_{\odot}}$\fi}

\shorttitle{H-atm NS Model Applied to X7}
\shortauthors{Heinke et al.}

\begin{document}
\title{A Hydrogen Atmosphere Spectral Model Applied to the Neutron
  Star X7 in the Globular Cluster 47 Tucanae}   

\author{Craig O. Heinke\altaffilmark{1,2}, George B. Rybicki, Ramesh Narayan,  Jonathan E. Grindlay}

\affil{Harvard-Smithsonian Center for Astrophysics,
60 Garden Street, Cambridge, MA  02138; 
 grybicki@cfa.harvard.edu,
 rnarayan@cfa.harvard.edu, jgrindlay@cfa.harvard.edu}

\altaffiltext{1}{Northwestern University, Dept. of Physics \&
  Astronomy, 2145 Sheridan Rd., Evanston, IL 60208; 
cheinke@northwestern.edu}

\altaffiltext{2}{Lindheimer Postdoctoral Fellow}


\begin{abstract}
Current X-ray missions are providing high-quality X-ray spectra from
neutron stars (NSs) in quiescent low-mass X-ray binaries (qLMXBs).
This has motivated us to calculate new 
hydrogen-atmosphere models, including opacity due to free-free
absorption and Thomson scattering, thermal electron conduction, and
self-irradiation by photons from the compact object.  We have
constructed a self-consistent grid of neutron star models covering 
a wide range of 
surface gravities as well as effective temperatures, which we make
available to the scientific community.

We present multi-epoch \Chandra X-ray observations of the qLMXB X7 in
the globular cluster 47 Tuc, which is remarkably nonvariable on
timescales from minutes to years.  Its high-quality X-ray spectrum is
adequately 
fit by our hydrogen-atmosphere model without any hard power-law
component or narrow spectral features.  If a mass of 1.4 \Msun\ is
assumed, our spectral fits require that its radius be in the range
$R_{\rm ns}=14.5^{+1.8}_{-1.6}$ km (90\% confidence), larger than expected
from currently 
preferred models of NS interiors.  If its radius is assumed to be 10 km,
then a mass of $M_{\rm ns}=2.20^{+0.03}_{-0.16}$ \Msun\ is required.   
Using models with the appropriate surface gravity for each value of the 
mass and radius becomes important for interpretation of the highest 
quality data.

\end{abstract}


\maketitle

\section{Introduction}\label{s:intro}

One of the primary goals of neutron star (NS) studies is to constrain the
behavior of matter at high densities by measuring NS masses and
radii \citep{Lattimer01}.  Mass measurements of high accuracy for a
number of radio 
pulsars in close binary systems (often with inferred NS companions)
are consistent with a range of NS masses between 1.25 and 1.45 \Msun
\citep{Thorsett99}.  
Fundamental constraints on NS interior structure can be achieved by
measurement of the gravitational redshift from the NS surface
\citep{Cottam02}.  Finally, it should be possible to derive
constraints on the radius of NSs from spectral fits to their X-ray
emission if the temperature, composition of atmosphere, and distance
to a NS are known, and the magnetic field is sufficiently weak so as
not to affect the opacity, or temperature distribution, on the NS
surface.  These requirements can be fulfilled for X-ray observations
of quiescent low-mass X-ray binaries (qLMXBs) containing NSs,
particularly those located in globular clusters where the distance is
well-known \citep{Brown98, Rutledge02a}. In this paper, we perform the
most accurate such test currently possible, using accurate hydrogen
atmosphere models constructed specifically for this project, and a
long \Chandra observation of a globular cluster containing a
relatively bright and remarkably constant qLMXB.

Several low-mass X-ray binaries which have been identified during
outbursts as accreting NS systems have been observed in quiescence
\citep[see][]{Campana98a, Rutledge02b}.  Their quiescent appearance
generally differs from that of quiescent systems containing black
holes both in their X-ray luminosity \citep{Garcia01} and their
observed X-ray spectrum \citep{Rutledge99,McClintock04}, both of which
indicate the presence of a compact object surface for NS systems and
not for black hole systems.  These NS systems generally show soft
spectra, consisting of a thermal, blackbody-like component, and
possibly a harder component extending to higher energies, usually fit
with a power-law of photon index 1-2 \citep[although some systems are
dominated by the harder component; ][]{Campana02,Wijnands05b}.  The
thermal component, if fit by a blackbody, produces inferred radii too
small for theoretical NS size estimates.  However, an accreting NS
will develop a pure hydrogen atmosphere if the accretion rate falls
below $\sim 10^{-13}$ \Msun yr$^{-1}$ \citep{Brown98}, because the
metals settle out of the atmosphere within a few seconds
\citep{Romani87}.  \citet{Rajagopal96} and \citet{Zavlin96} showed
that hydrogen NS atmospheres shift the peak of the emitted radiation
to higher frequencies due to the strong frequency dependence of
free-free absorption.

A major unsolved question about qLMXBs is the nature of the
X-ray emission.  \citet{Brown98} advanced the idea \citep[also
  discussed by][]{Campana98a} that the soft thermal component seen in
these field 
systems can be explained by the release, over long timescales, of heat
injected into the deep crust by pycnonuclear reactions driven during
accretion (the ``deep crustal heating'' model).  This scenario
generally predicts the quiescent thermal luminosity of many qLMXBs,
based on their 
outburst history, reasonably well \citep{Rutledge01a}, but not for all
qLMXBs \citep[cf.][]{Colpi01, Campana02, Wijnands05b}. 
The deep crustal heating model cannot explain the hard power-law
component, which is often attributed to continued accretion and/or a
shock from a pulsar wind \citep{Campana98a,Bogdanov05}.  
  The deep crustal heating
model also cannot explain the short-timescale ($\sim10^4$ s) variability
observed from Aquila X-1 \citep{Rutledge02b} and Cen X-4
\citep{Campana04}.  Continued accretion has been suggested as an
explanation for the thermal component, as the radiation spectrum from
matter accreting radially 
onto a neutron star should be similar to that expected from deep
crustal heating \citep{Zampieri95}.  If an absorption feature due to
metals in the NS atmosphere 
were to be confirmed in a qLMXB spectrum \citep[as suggested
  in][]{Rutledge02b}, this would provide evidence 
for continued accretion at rates sufficient to explain most or all of
the thermal emission.

Applying the hydrogen-atmosphere NS 
 models of \citet{Zavlin96} to \Chandra  observations of
 qLMXBs,
 \citet{Rutledge99, Rutledge01b, Rutledge01a} have shown that the 
 radius predictions of the models are consistent with the range of
 radii expected from NSs.  These analyses have suffered from 
 uncertainties in the distances to qLMXB systems, and from
 uncertainties due to fitting two components (the 
 power-law plus thermal components) to a spectrum.  The distance is
tightly constrained for some globular clusters \citep[4\% distance 
uncertainty to 47 Tuc,][]{Gratton03}, making them an excellent target
 for such studies, as pioneered by \citet{Rutledge02a,Gendre03a,
 Heinke03a}.  The well-studied globular cluster 47  
Tucanae (NGC 104; hereafter 47 Tuc) is especially ideal due to its
close distance (4.85$\pm0.18$ kpc), low
reddening \citep[$E(B-V)=0.024\pm0.004$,][]{Gratton03}, presence of
 two reasonably bright qLMXBs \citep[X5 and X7;][hereafter
 HGL03]{Heinke03a}, and deep \Chandra observations \citep[300
 ksec,][]{Heinke05a}. 

To match the quality of the best \Chandra data we desire highly
accurate hydrogen-atmosphere NS models.  
We have been troubled by the disagreement between the predictions of
the currently available models of \citet{Zavlin96} and
\citet{Gansicke02}.  We also wished to verify whether the variation in
surface gravity over the relevant range in NS mass and radius has a
significant effect on the atmosphere models and spectral fitting.  For
these reasons we have produced grids of new hydrogen-atmosphere
models.  We consider only models of pure hydrogen and for which the
magnetic field is sufficiently weak ($B \lesssim 10^8$ G) that it may
be ignored in determining the spectrum. The former assumption is
consistent with previous observations of qLMXBs, as iron or solar-abundance
atmospheres would be easily identifiable by their different spectral
shapes (however, subtle departures from pure H might still go
unnoticed, and affect our results). The latter assumption is
consistent with the 
lack of observable millisecond time variability in qLMXBs similar
to X7 \citep[e.g. Aql X-1,][]{Chandler00}; the accreting millisecond X-ray
pulsars appear to be substantially fainter and harder in quiescence
\citep{Wijnands05a}.  

A number of codes exist to compute the spectrum for the simple case of
a pure hydrogen NS atmosphere with no magnetic field.  Some of these
are limited to zero magnetic field, e.g., \citet{Rajagopal96},
\citet{Gansicke02}, and \citet{McClintock04}.  Others include
magnetic fields, but can be applied in the zero field limit, e.g.,
\citet{Zavlin96} and \citet{Lloyd03}.  

Even with these simplifying assumptions, grids of NS atmosphere models
over a wide range of parameters are not widely available for use in
XSPEC.  Some researchers have computed their own model atmospheres to
be used in XSPEC and have included ranges of gravity as well as
effective temperature, e.g., \citet{Zavlin98}, using NSA, and
\citet{Stage04}, using ATM.  However, the available models included
in the standard XSPEC package, namely NSA($g_s$) \citep{Zavlin96} and
HYD\_SPECTRA($g_s$) \citep{Gansicke02} cover a range of effective
temperatures, but only for the single surface gravity $\log g_s =
14.385$, corresponding to a ``standard'' NS model with mass $M_{\rm
ns} = 1.4 M_{\odot}$ and radius $R_{\rm ns} = 10$ km [we shall emphasize
this using the qualifier ``($g_s$)''].  However, as the
results of this paper will demonstrate, the surface gravity can play
an important role in the spectral fitting and needs to be taken into
account.

A large part of the present effort was devoted to the construction of
grids of models to fulfill this need.  The need for NS models covering
an extensive set of parameters, including surface gravity, motivated us
to adapt the code NSATMOS.  This code was previously used by
\citet{McClintock04} to investigate the hydrogen atmospheres of
hypothetical, compact objects, some so compact that they lay within
their own photon spheres; this required taking account of the
self-irradiation of the surface due to gravitational bending of rays.
The same model assumptions described in the Appendix of McClintock et
al.\ (2004) 
apply here, with some improvements: the code now includes the
radiation force in the equation of hydrostatic equilibrium 
\citep[e.g. ][]{Pavlov91}, and
includes heat conduction by electrons in the energy equation 
\citep[e.g. ][]{Rajagopal96}.
These and other technical improvements made to NSATMOS are discussed
in Appendix A.  

As it turned out for the present problem, neither electron conduction
nor radiation force played any substantial role, in agreement with the 
results of, e.g., \citet{Gansicke02}.  Also, the range of
parameters where self-irradiation occurs does not overlap with the range 
allowing neutron stars which obey causality, indicating self-irradiation 
is unlikely to be relevant for neutron stars.   On the other hand, allowing for
variations in surface gravity turned out to be very important. A major
advantage of NSATMOS for this problem was its 
speed, which allowed us to compute extensive grids of models to cover
the ranges of effective temperature and gravity tailored to our needs.
Our NSATMOS code will be made available to the astronomical community
through the XSPEC
website\footnote{http://heasarc.gsfc.nasa.gov/docs/xanadu/xspec/}. 

 Our \Chandra observations are described in \S \ref{s:obs}. We compare
the models and data in \S \ref{s:spec}, and discuss the implications
in \S \ref{s:disc}.  Our new neutron star atmosphere
models are described in detail in Appendix A, and the effects
of our consideration of varying surface gravity are discussed in
Appendix B.


\section{Observations}\label{s:obs}

The data used in this paper are from the 2000 and 2002 \Chandra
observations of the globular cluster 47 Tuc.  Both sets of
observations and their initial reduction are described in detail in
\citet{Heinke05a}; prior analyses of the 2000 dataset are described in
\citet{Grindlay01a} and HGL03.  The 2000 observations were performed
with the ACIS-I CCD array at the telescope focus, while the 2002
observations placed the back-illuminated ACIS-S aimpoint at the focus
for maximum low-energy sensitivity.  Five consecutive observations
were performed in 2000, as listed in Table 1, 
\placetable{tab:obs}
with three short observations, using a subarray and faster readout
time, interleaved to reduce pileup in bright sources such as X7 and
X5 (see HGL03). Pileup occurs when two X-ray photons, arriving at the detector
during one frame time, are erroneously identified as a single photon
with the sum of the two photon energies, or else discarded
\citep[see][]{Davis01}. Pileup has effects upon both spectral and
timing analyses, as discussed in the Chandra Proposers' Observatory 
Guide\footnote{http://cxc.harvard.edu/proposer/POG/} and \citet{Davis01}. 

\begin{deluxetable}{lccccr}
\tabletypesize{\footnotesize}
\tablewidth{4.8truein}
\tablecaption{\textbf{Summary of \Chandra Observations}}
\tablehead{
\colhead{Seq, OBSID} & \colhead{Start Time} & \colhead{Exposure} &
\colhead{Aimpoint} & \colhead{Frametime} & \colhead {CCDs}
}
\startdata
300003,078 & 2000 Mar 16 07:18:30  &  3875  & ACIS-I & 0.94 & 1/4 \\
300028,953 & 2000 Mar 16 08:39:44  & 31421  & ACIS-I & 3.24 & 6 \\
300029,954 & 2000 Mar 16 18:03:03  &   845  & ACIS-I & 0.54 & 1/8 \\
300030,955 & 2000 Mar 16 18:33:03  & 31354  & ACIS-I & 3.24 & 6 \\
300031,956 & 2000 Mar 17 03:56:23  &  4656  & ACIS-I & 0.94 & 1/4 \\
400215,2735 & 2002 Sep 29 16:59:00 & 65237 & ACIS-S & 3.14 & 5 \\
400215,3384 & 2002 Sep 30 11:38:22 &  5307 & ACIS-S & 0.84 & 1/4 \\
400216,2736 & 2002 Sep 30 13:25:32 & 65243 & ACIS-S & 3.14 & 5 \\
400216,3385 & 2002 Oct 01 08:13:32 &  5307 & ACIS-S & 0.84 & 1/4 \\
400217,2737 & 2002 Oct 02 18:51:10 & 65243 & ACIS-S & 3.14 & 5 \\
400217,3386 & 2002 Oct 03 13:38:21 &  5545 & ACIS-S & 0.84 & 1/4 \\
400218,2738 & 2002 Oct 11 01:42:59 & 68771 & ACIS-S & 3.14 & 5 \\
400218,3387 & 2002 Oct 11 21:23:12 &  5735 & ACIS-S & 0.84 & 1/4 \\
\enddata
\tablecomments{Times in seconds. Subarrays are indicated by fractional
  numbers of CCDs.  We do not use OBS\_ID 3385 in this paper, due to
  its relatively high background. \label{tab:obs} }
\end{deluxetable}

 The 2002
observations also interleaved long observations with shorter ones to
collect some data relatively free of pileup.  The countrates for X7 in
the 2000 and 2002 observations (both essentially on-axis) were 0.07
and 0.12 cts/s, leading to predicted pileup rates of 9\% and 15\% for
full-array observations, or 2\% and 4\% using 1/4 subarrays.  We
reprocessed the 2000 observations using the CTI correction algorithm
implemented in CIAO 3.2 acis\_process\_events.  Both the 2000 and 2002
observations were reprocessed (using CIAO 3.2) to remove the 0\farcs5
pixel randomization added in standard processing, use updated
(time-dependent) gain files, and improve hot pixel identifications.
Some background flaring occurred during parts of the 2002
observations, particularly affecting OBS\_ID 3385.  We do not use data
from that short (5 ksec) observation, but keep all other data.

We used the ACIS\_EXTRACT software \citep{Broos02}, version 3.65, to
extract and combine spectra and response files for the various
observations.  We extracted spectra from contours matching the 95\%
encircled energy (at 1.5 keV) of the \Chandra point-spread function at
X7's position.  We constructed response files using the {\it
  mkacisrmf} response generator in CIAO 3.2.  The effective area files
take into account the 
decreasing quantum efficiency of the ACIS chips, and are corrected to
account for the energy-dependent fraction of the point-spread function
enclosed by the extraction region \citep{Broos02}.  
We combined spectra from: the two long 2000 observations, OBS\_IDs 953
and 955; the four long 2002 observations; the three remaining short 2002
observations (see above); and the three short 2000 observations, for a
total of four summed spectra.  We
binned these spectra at 80 counts/bin for the long 2002 spectrum, 20
counts/bin for the short 2000 spectrum, and 40 counts/bin for the
other two spectra.  
\clearpage

\subsection{Timing Analysis}\label{s:timing}

We adjusted all event times to the solar system barycenter using
satellite orbit files provided by the \Chandra X-ray Center.  
The qLMXB X5 continues to show eclipses and very strong dipping
activity throughout the 2002 observations, which make a study of its
spectrum more complicated; we defer detailed studies of X5 to a later
paper. 

Kolmogorov-Smirnov and Cramer-von Mises tests showed no variability
from X7 on any timescales probed by the 2002 observations (seconds to
weeks).  Power spectra (constructed using XRONOS) of X7's lightcurves
showed a flat power spectrum with less power than expected from
Poisson processes \citep{Leahy83}, typically $\sim70$\% of the
expected level.  We 
attribute this to the effects of moderate pileup (as 
other, weaker, X-ray sources in 47 Tuc showed the expected levels of
white noise), and
note that it makes a quantitative limit on X7's variability difficult
to determine.  Assuming constant Poisson noise (the constant level was
a free parameter), plus red noise with a
fixed slope \citep[$\propto \nu^{-1}$, e.g.][]{Rutledge01b}, we 
constrain ($3\sigma$, between $10^{-5}$ and 0.1 Hz) the rms variability
to $<3.6$\%, $<7.3$\%, $<1.9$\% and $<5.2$\% for the four 2002 
observations. These values may be underestimates due to the effects of
pileup.  We also extract power spectra from 1-2\farcs annuli, to
reduce the effects of pileup, giving 511
to 541 counts per dataset.  These power spectra display white noise at
the levels expected from Poisson noise.  We again find no evidence for excess
variability, with $3\sigma$ upper limits on rms variability being 
$<$17\%, $<$15\%, $<$25\% and $<$28\% for each observation. 
The data taken in subarray mode to reduce pileup showed similar
properties, with $3\sigma$ upper limits of $<$13\%, $<$17\%, and
$<$24\% rms variability (excluding OBS\_ID 3385).
 No signal was seen at 5.50 hours, the period of a 
marginal signal identified in HGL03, indicating that the suggested  
orbital period (from the lower-quality 2000 data) is probably spurious.

\section{Spectral Analysis}\label{s:spec}

\subsection{NSATMOS Spectral Analysis of X7}

Our standard XSPEC model consists of the NSATMOS hydrogen atmosphere
model, absorbed 
by interstellar gas, convolved with the XSPEC pileup model. 
We use the XSPEC \citep{Arnaud96} version of the
pileup model of J. Davis \citep{Davis01}, setting 
the frame time parameter to the time resolution of each 
spectrum (header keyword TIMEDEL), and the
parameters $g_{0}$=1 and $psffrac$=0.95. We floated the grade
migration parameter $\alpha$ (which parametrizes the fraction of
piled photons that are recorded as good grades), but found
that its value is always consistent with 0.5, and fixed it to 0.5 for
 this section and \S \ref{s:edgepl}. 

We use the XSPEC {\it phabs} model, with \citet{Wilms00}
interstellar element abundances, to describe the interstellar  
gas between the Earth and 47 Tuc. 
We fix its normalization at
$N_H=1.3\times10^{20}$ cm$^{-2}$, derived using E(B-V)=0.024$\pm0.004$  as 
measured by \citet{Gratton03}, and assuming
$R_V$=3.1 \citep{Cardelli89} and $N_H/A_V=1.79\times10^{21}$ cm$^{-2}$
\citep{Predehl95}.  The $N_H$ column measured using any of our models
is larger than this, indicating additional gas intrinsic to the system (or 
slight errors in the \Chandra ACIS calibration below 1 keV).
We model this additional absorbing gas by the XSPEC {\it vphabs}
model, selecting the abundances 
to correspond to those of 47 Tuc.  We choose the abundance of
iron to be 20\% of solar ([Fe/H]=-0.7), that of metals between Ne and
Ca to be 40\% solar ([X/H]=-0.4), C, N, and O to be 63\% solar
([X/H]=-0.2), and He at solar abundance \citep[using][]{Carney96,
  Salaris98, Gratton03}.   We note that
our results are not very sensitive to the
detailed abundances, since the absorption column is low (see below).

The \Chandra ACIS effective area and response matrix calibrations are
very uncertain below 0.5 keV.  We find substantial negative residuals
around 0.4 keV, similar to an absorption line (but substantially
narrower than the instrument resolution).  This feature is seen in
most other X-ray sources of sufficient flux in 47 Tuc
\citep{Heinke05a}, so we ascribe 
it to an instrumental effect and ignore data below 0.5 keV in this
paper.  A second feature is a wave in the 
residuals between 1.7 and 2.3 keV.  This feature can be ascribed to
the difficulty of calibrating the iridium M edge structure (Chandra
Proposer's Observatory Guide), and similar waves are seen in other
high-quality ACIS CCD spectra \citep{Sanders03}.  This residual causes
 the quality of our best fits to be slightly less than nominal, 
$ \chi_{\nu}^2$=1.20 for a null hypothesis probability (nhp) of 1.5\%.
More acceptable fits (nhp=7\%) can be achieved by adding a systematic error
term (of order 2\%), or by excluding the spectral range from 1.9-2.25
keV in the two highest-quality spectra, but the parameters of the fit
do not change substantially from the fits using all the data.

\vspace*{0.3cm}
\begin{figure}
\figurenum{1}
\includegraphics[angle=270,scale=.36]{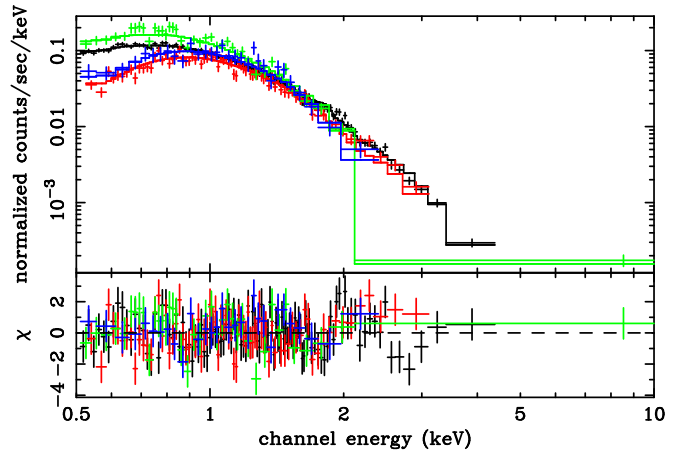}  
\caption[X7_vph_nsat_data.ps]{ \label{fig:data}
From top, \Chandra ACIS-S subarray spectrum (green), ACIS-S full-frame
spectrum (black), ACIS-I subarray spectrum (blue), and ACIS-I
full-frame spectrum (red) of
X7, fit with our hydrogen-atmosphere NSATMOS model (histograms) and
photoelectric absorption (see text).  Pileup is responsible for the
variation in count rate between the subarray and full-frame
spectra,and is accounted for in the fit. The apparent difference between
the full-frame spectra is due to the differing spectral responses of
the ACIS-S and ACIS-I instruments, as no model parameter is allowed to
vary between the fits.  The strongest residuals (near 2 keV) are
instrumental features due to the iridium edges of the mirror and
resultant rapid changes in effective area with energy.  
 {\it See the
  electronic edition of the Journal for a color version of this figure.}
} 
\end{figure}

Fitting the four combined X7 spectra simultaneously in XSPEC, leaving
no parameters free between the various datasets, and only the
intrinsic absorption, NS temperature, and NS radius as free
parameters, gives a reasonably good fit ($\chi^2_{\nu}$=1.20,
nhp=1.5\%; see Fig.~\ref{fig:data}).  \placefigure{fig:data} (To begin
with, we fix the NS mass at 1.4 \Msun.  We will vary this later.)  The
inferred (unabsorbed) luminosity of X7 is $L_X$(0.5-10
keV)$=1.5\times10^{33}$ ergs s$^{-1}$, $L_{\rm bol}=2.4\times10^{33}$
ergs s$^{-1}$.  Allowing the absorption column or the NS temperature,
radius, normalization, or distance to vary between the 2002 and 2000
observations does not improve the fit, and the best-fit values for
each observation lie within the one-sigma errors of the best fit for
the other observation.  A quantitative constraint on X7's spectral
variability between the 2000 and 2002 observations can be obtained by
decoupling the 2000 and 2002 temperatures or normalizations, and
measuring how large the difference may be; we find $kT_{X7, 2000}=
1.003^{+0.006}_{-0.006} \times kT_{X7, 2002}$, or for normalization,
$K_{X7, 2000}=1.01^{+0.03}_{-0.03} \times K_{X7, 2002}$ (90\%
conf.). Fitting the individual spectra extracted for each of the four
long 2002 observations (grouped at 20 counts/bin) with the same model
gives a good fit ($\chi^2_{\nu}$=0.962, nhp=70\%), and when the model
parameters are allowed to vary between observations they again agree
with one another. (The improvement in the fit quality, compared to the
combined spectrum, is due to the reduced statistics of each
observation individually.)  Therefore we conclude that these deep
observations provide no evidence for any spectral change in X7 over a
period of 2.5 years.  This result is in contrast to other qLMXBs which
have displayed substantial variability in quiescence
\citep{Rutledge02b, Campana04}.  

\subsection{Edge or power-law?}\label{s:edgepl}

HGL03 found marginal evidence for the existence of an edge
  or other absorption feature near 0.63 and 0.66 keV in X7 and X5,
  respectively.  No such feature is
apparent in the 2002 X7 data, and the evidence for such a feature from
the 2000 data has decreased as the \Chandra calibration has improved.
Fitting all the X7 data with the standard model above and an edge fixed at
0.63 keV (as in HGL03) does not improve the fit, and leads to a 90\%
  confidence upper limit of $\tau<0.045$ for the edge.   The two
  explanations for the edge suggested by HGL03 (an O\textsc{V} edge
  from an ionized wind,  
or a signature of a NS atmosphere that is not purely hydrogen) can
both be excluded by the disappearance of this feature in the 2002
dataset.  We feel that the most likely explanation for this apparent
feature is small uncertainties in the calibration.  

The existence of any spectral feature in a qLMXB spectrum would be of
great interest, since an identification of the feature could allow
determination of the gravitational redshift at the NS surface
\citep[][HGL03]{Brown98,Rutledge02b}.  Therefore we searched for any
possible features in X7's spectrum, using an edge or a gaussian
absorption line, between 0.55 and 3 keV.  
The largest possible features were at
0.85 keV, where an edge with $\tau=0.065^{+0.07}_{-0.05}$ gave $\chi^2_{\nu} =
1.017$; 1.58 keV, where an edge with $\tau=0.14^{+0.11}_{-0.10}$ gave
$\chi^2_{\nu} = 1.194$; and 2.54 keV, where an edge with
$\tau=0.70^{+1.9}_{-0.7}$ gave $\chi^2_{\nu} = 1.204$. The latter two
are probably caused by the calibration uncertainties around 2 keV.
These features have F-test probabilities (to be interpreted with caution) of
7.7\%, 13\%, and 45\% of being  
generated by chance, which are consistent with the low significances of their
optical depths (none nonzero at more than 98\% confidence).  The
90\% confidence limits on edge equivalent widths at these locations are
$<18$, $<29$, and $<53$ eV.  Using a gaussian absorption line with
fixed intrinsic width of 0.05 keV, we find similar results, with the
most likely features located at 0.92 ($\tau=0.083^{+0.05}_{-0.05}$) or
1.73 ($\tau=0.33^{+0.20}_{-0.15}$) keV.  Equivalent width upper limits
are $<11$ and $<33$ eV, respectively.  

The feature at $\sim$0.4 keV discussed above is the strongest apparent 
feature.  The anticipated energy of the strongest feature (due
principally to the O \textsc{VIII} edge) likely to appear 
on an accreting neutron star with a near-solar abundance atmosphere is
$0.87/(1+z)$ keV  
\citep[see their Fig.\ 1]{Brown98,Rutledge02b}.  There are no 
plausible features expected near 0.4 keV for gravitational redshifts in the
range 0.2-0.4 (typical for favored neutron star equations of state).
We conclude that no spectral features 
intrinsic to the NS atmosphere have been detected.  

The hard power-law component identified in many field qLMXBs may be a 
 signal of continued accretion \citep{Rutledge02b}, or of nonthermal 
emission from the pulsar wind of an underlying MSP 
\citep{Burderi03, Bogdanov05}. We
constrain such a power-law component 
by adding it to our standard model with a fixed photon index $\Gamma$ of 1.5,
2 or 1, and deriving constraints upon its flux in the 0.5-10 keV band
relative to the total (absorbed) 0.5-10 keV flux.  Adding this
component does not significantly improve the fit ($\chi^2_{\nu} =
1.205$, an F-test suggests a 60\% chance of this level of improvement
by chance).  For $\Gamma=1.5$, such a
power-law component would make up $<3.2$\% (90\% conf.) of X7's 
total 0.5-10 keV flux.   For $\Gamma=2$ or 1, the constraints are
$<3.4$\% of the total 0.5-10 keV flux.  

The complete lack of evidence for variability, edges, or hard power-law 
components in X7 suggests that X7's X-ray emission is produced entirely 
through re-emission of stored heat \citep{Brown98}.  The constraints on 
these properties for X7, more stringent than for any other qLMXB, 
make this an ideal object for comparisons with NS hydrogen-atmosphere 
models.

\subsection{Constraining the NS Mass and Radius}\label{s:RM}

The principal goal of our spectral fitting is to self-consistently
constrain the allowed space in mass and radius of X7's neutron star.  
First we explore the best fits for mass and radius if one parameter is
fixed at a canonical value, $M_{\rm ns}$=1.4 \Msun\ or $R_{\rm ns}$=10
km.  Our standard model above  
keeps the mass fixed at 1.4 \Msun, and the fitted radius is
14.2$^{+1.1}_{-1.0}$ km (90\% conf.).  Allowing the pileup parameter
$\alpha$ to vary, and allowing for a power-law component with photon
index fixed at 1.5, the fitted radius is 14.5$^{+1.6}_{-1.4}$ km.  The
parameters for this fit are included in Table~2.  
\placetable{tab:spec}
We note
that a pure helium atmosphere (possible if the donor is degenerate) would
give even larger radii \citep{Zavlin96,Pons02}, which we think
unlikely.  
 Adding a 6.1\% uncertainty \citep[][90\% conf., assuming errors are
normally distributed]{Gratton03} in the distance to 47 Tuc gives a
 NS radius of 14.5$^{+1.8}_{-1.6}$ km.  The lower limit to X7's 
 radius is thus 12.9 km, which is significantly larger than the radius
 predicted by the APR \citep[11.5 km,][]{Akmal98} and FPS 
\citep[10.8 km,][]{FPS89} equations of state for a 1.4 \Msun\ NS.  

\begin{deluxetable}{ccc}
\tablewidth{3.2 truein}
\tablecaption{\textbf{X7 Spectral Model Parameters}}
\tablehead{
\colhead{\textbf{Model Parameter}} & \colhead{\textbf{M fixed}} &
\colhead{\textbf{R fixed}} 
}
\startdata
\hline
\multicolumn{3}{c}{\textbf{Rybicki NSATMOS model} } \\ 
\hline
$kT$, eV & $105.4^{+5.6}_{-5.6}$ & $151.5^{+7.4}_{-5.9}$ \\
$N_{H,20}^a$ & $4.2^{+1.8}_{-1.6}$ & $4.0^{+1.8}_{-1.5}$ \\
$R_{\rm ns}$, km & $14.5^{+1.6}_{-1.4}$ & (10.0) \\
$M_{\rm ns}$ & (1.4) & $2.20^{+0.16}_{-0.15}$  \\
$\chi^2_{\nu}$/dof & 1.21/251 &  1.21/251 \\
Null hyp. prob. & 1.3\% & 1.3\%\\
\hline
\multicolumn{3}{c}{\textbf{Zavlin NSA($g_s$) model} } \\ 
\hline
$kT$, eV &  $89^{+5}_{-1}$ &  $145^{+4}_{-5}$  \\
$N_{H,20}^a$ &   $5.6^{+0.8}_{-0.4}$ &  $3.5^{+5.0}_{-2.8}$  \\
$R_{\rm ns}$, km &  $19.9^{+0.1^b}_{-2.1}$ &  (10.0) \\
$M_{\rm ns}$ &  (1.4) &  $1.99^{+0.13}_{-0.15}$  \\
$\chi^2_{\nu}$/dof & 1.21/251  & 1.21/251  \\
Null hyp. prob. &  1.2\% &  1.3\% \\
\hline
\multicolumn{3}{c}{\textbf{G\"{a}nsicke HYD\_SPECTRA($g_s$) model} } \\ 
\hline
\colhead{-} & \colhead{\textbf{z = 0.306}} &
\colhead{\textbf{R fixed}} \\
\hline
$kT$, eV &  $123^{+4}_{-5}$ &  $138^{+5}_{-4}$  \\
$N_{H,20}^a$ &   $5.1^{+0.8}_{-0.9}$ &  $4.4^{+1.6}_{-0.8}$  \\
$R_{\rm ns}$, km &  $11.9^{+1.5}_{-1.2}$ &  (10.0) \\
$M_{\rm ns}$ &  $1.66^{+0.22}_{-0.15} $ &  $1.68^{+0.20}_{-0.17}$  \\
$\chi^2_{\nu}$/dof & 1.21/251  & 1.21/251  \\
Null hyp. prob. &  1.3\% &  1.3\% \\
\enddata
\tablecomments{ All errors are 90\% confidence limits.  Distance of
4.85 kpc is assumed (its uncertainty is not included in these quoted
errors). In each column, either mass, redshift ($z$), or 
the true radius of the neutron star is held fixed.
$^a$ $N_H$ intrinsic to system, in units of 10$^{20}$ cm$^{-2}$, in
addition to galactic column of 1.3$\times10^{20}$ cm$^{-2}$.
$^b$ Reached hard limit of model.  \label{tab:spec}
}
\end{deluxetable}

If we instead allow $M_{\rm ns}$ to vary and fix $R_{\rm ns}$ at 10
km, we find that $M_{\rm ns}$ is 
constrained to 2.17$^{+0.10}_{-0.12}$ \Msun; allowing $\alpha$ to vary
and including a possible powerlaw component,
$M_{\rm ns}$=2.20$^{+0.16}_{-0.15}$ \Msun.  Adding distance uncertainty, the errors
are $M_{\rm ns}$=2.20$^{+0.18}_{-0.16}$ \Msun.  Constraining the neutron star mass 
to lie below the causality line gives $M_{\rm ns}$=2.20$^{+0.03}_{-0.16}$ 
\Msun.
 This mass is significantly
larger than that of any well-measured NS \citep[but cf. ][]{Nice05}.
  A high mass for X7 seems improbable
given its relatively high X-ray luminosity (compared to other qLMXBs),
and the expected tendency for 
more massive NSs to cool more quickly \citep{Yakovlev04}.  We give the
parameters for fits with either $M_{\rm ns}$ or $R_{\rm ns}$ held fixed in
Table~2, including 
all uncertainties except for the 6\% distance uncertainty. 

\begin{figure}
\figurenum{2}
\includegraphics[angle=0,scale=.42]{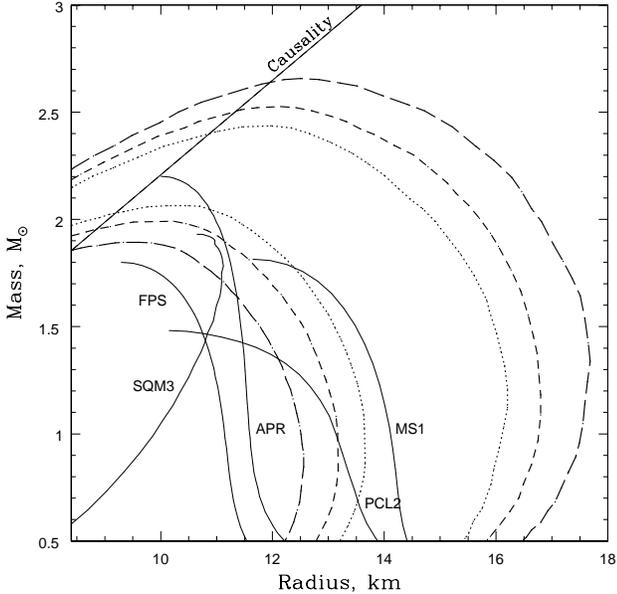}
\caption[X7rybcont.eps]{
68\% (dotted), 90\% (short dash) and 99\% (long dash) confidence
contours in the mass-radius plane 
derived for X7 by our spectral fitting with the NSATMOS model.
The causality line, above which no realistic NS equations of state can
exist, is plotted, along with five representative equations of state, 
APR, FPS, MS1, PCL2, and SQM3 \citep[the last a quark star model,
  see][]{Lattimer01}. 
\label{fig:rybcont}}
\end{figure}

  We use the {\it steppar} command in XSPEC to vary both the
radius and mass  parameters, allowing all
other free parameters (including $\alpha$ and the powerlaw
normalization) to vary to find the best fit.   We show
1$\sigma$, 90\% confidence, and 99\% confidence ($\Delta\chi^2=2.3$, 4.61, and
9.21 respectively) contours in NS mass and radius in
Fig.~\ref{fig:rybcont} for X7, using our spectral model as 
\placefigure{fig:rybcont} 
described above.  These can be compared with the loci of models
following the APR \citep{Akmal98}, FPS \citep{FPS89},  PCL2 
\citep{Prakash95}, MS1 \citep{Muller96}, and SQM3 
\citep{Prakash95} equations of state for dense matter (the last is a
representative ``quark star'' model).  For the SQM3 and APR models,
a high NS mass is required for consistency with the data; the FPS
model is ruled out; and the PCL2 model is marginally consistent with
the data at the 99\% level for a mass of 1.35 \Msun.   The 
MS1 model is an example of models consistent with a larger radius at
1.4 \Msun\ \citep[see][]{Lattimer01}, some of which include
condensation of kaons, hyperons, muons or free quarks in the core.  
The line labeled ``Causality'' ($R=3.04 G M /c^2$) represents the
requirement that the speed of sound must be less than $c$, a necessary
(but not sufficient) requirement for a NS interior to respect
causality \citep{Lattimer01,Olson01}.  This requirement is virtually
  identical to $R>3GM/c^2$, which is the condition that the neutron
  star surface is outside its own photon sphere, so that
  self-irradiation does not occur.

\subsection{Comparison of NS atmosphere models}\label{s:othermod}

We compare three different hydrogen-atmosphere neutron star models in this
analysis.  Our NSATMOS model has parameters $M_{\rm ns}$ (mass of neutron star,
\Msun), $R_{\rm ns}$ (true radius of the neutron star, km), $\Teff$ 
(the effective temperature of the neutron star surface, unredshifted),
and $D$ (distance to neutron star, pc).
The NSA model \citep{Zavlin96} has parameters $R$ (true radius of
the neutron star, km), $M$ (mass of neutron star, \Msun), 
$\Teff$ 
(the effective temperature of the neutron star surface, unredshifted),
and normalization $K=1/D^2$ (where $D$ is the distance to the neutron
star, pc).  The last parameter is often 
used to calculate $R_{\infty}=R (1+z)$, the radius as seen by a distant
observer, since the last parameter can be considered
$K=(R_{\infty}/R_{\infty,0})^2 / D^2$, where $R_{\infty,0}$ is the
equivalent $R_{\infty}$ that would be calculated from the model
parameters $R$ and $M$.  
The HYD\_SPECTRA model of \citet{Gansicke02} has the parameters
$\Teff$, $z$ (the gravitational redshift at the NS surface), and
normalization $K=(R/D)^2$, where $R$ is in units of 10 km and $D$ is
measured in pc. 
We also tested the H\_ATM model of \citet{Lloyd03}, for which the
parameters are $\Teff$, $z$, and normalization $K=(R/D)^2$, where $R$ is in km
and $D$ is in units of 10 kpc.  The H\_ATM model gives results broadly
similar to those of the NSA model, and for brevity we will not
discuss H\_ATM further in this paper.
The NSATMOS and NSA models fit directly for $M_{\rm ns}$ and $R_{\rm
  NS}$, while the 
HYD\_SPECTRA model fits redshift and normalization, which allows computation
of $M_{\rm ns}$ and $R_{\rm ns}$.  

  While the NSA and HYD\_SPECTRA codes can, in principle, be used to
compute models for any value of gravity, the actual versions of these 
models found in the standard XSPEC package have been constructed for single
values of the surface gravity $g_s$ (appropriate for a 1.4 \Msun, 10
km NS).  To emphasize this, we shall denote these models by NSA($g_s$)
and HYD\_SPECTRA($g_s$).  The use of these fixed gravity models
for other values of surface gravity is not strictly appropriate.  In
order to account for the effect of surface gravity on the spectra, we
computed an extensive grid of models using the NSATMOS code described
in Appendix A.  This grid was used as the basis of an interpolation
routine that produced model spectra for the XSPEC program for a full
range of effective temperatures and gravities.

\begin{figure}
\figurenum{3}
\includegraphics[angle=0,scale=.42]{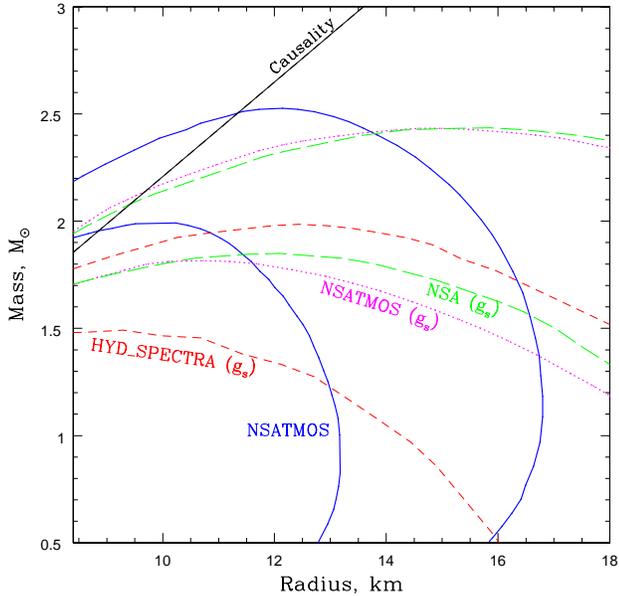}
\caption{
90\% confidence contours in the mass-radius plane derived for X7 by
our spectral fitting with hydrogen atmosphere NS models: NSATMOS, using
variable $g$, with thick solid (blue) lines, NSA($g_s$)
\citep{Zavlin96} with dashed (green) lines, HYD\_SPECTRA($g_s$)
\citep{Gansicke02} with short-dashed (red) lines, and, for comparison,
NSATMOS($g_s$), with thick dotted (magenta) lines.  Here $g_s=14.385$
is the surface gravity for the standard NS model.
{\it See the
  electronic edition of the Journal for a color version of this figure.}
\label{fig:allcont}}
\end{figure}

We fit each of these three models to our X7 data using the same
spectral model as described in section \S \ref{s:RM}, replacing only
the neutron-star atmosphere model.  For constraints upon the NS mass,
radius, or gravitational redshift, we place the parameters of these
fits in Table~2.  We plot the contours enclosing 90\% confidence contours for
various models together in Fig.~\ref{fig:allcont}.  The contours
labelled NSATMOS were constructed using the appropriate surface
gravity for each point in the plot.  We have also plotted the
contours assuming a fixed standard gravity $\log g_s=14.385$ for three codes: 
NSATMOS($g_s$), NSA($g_s$), and HYD\_SPECTRA($g_s$), the latter
two using the models in the XSPEC package.

We see that none of the available models allow a 1.4 \Msun, 10 km
radius NS, requiring a more massive or larger radius NS.  [However,
the small additional uncertainty on the distance to 47 Tuc--not
included in Fig.~\ref{fig:allcont}--allows marginal consistency at the
99\% level with the HYD\_SPECTRA model($g_s$).]  Since the
HYD\_SPECTRA($g_s$) and NSA($g_s$) models are constructed with a
$g_s$ appropriate for a 1.4 \Msun, 10 km NS, the failure of those
models to find an acceptable fit for a 1.4 \Msun, 10 km NS is robust.
It is not clear to us why the HYD\_SPECTRA($g_s$) model predictions
are significantly different from the other two models, although the
difference between the HYD\_SPECTRA and NSA hydrogen models has been
previously noted \citep{Gansicke02}.  Caution should still be taken 
in interpreting our results due to possible 
unknown systematic uncertainties in the ¡ÀChandra calibration.  However,
 the robustness of all tested models in excluding the canonical mass and 
radius values (1.4 \Msun and 10 km) is strong evidence for a relatively 
``stiff'' NS equation of state.

\section{Discussion}\label{s:disc}

\begin{figure}
\figurenum{4}
\includegraphics[angle=0,scale=.42]{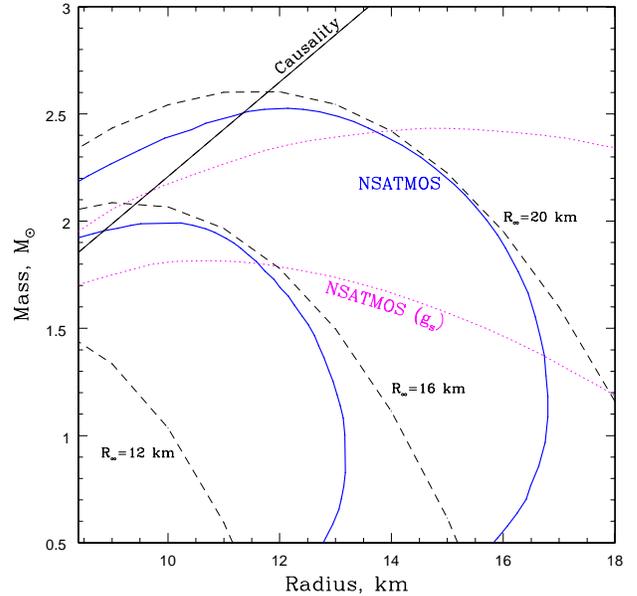}
\caption{
90\% confidence contours in the mass-radius plane
derived for X7 by our spectral fitting with two versions of our hydrogen
atmosphere NS models: NSATMOS (variable $g$) and NSATMOS($g_s$), 
compared to lines of constant $R_{\infty}$.  
  {\it See the
  electronic edition of the Journal for a color version of this figure.}
\label{fig:rinf}}
\end{figure}

In Fig.~\ref{fig:rinf}, we compare the allowed contours of mass and
\placefigure{fig:rinf} radius for X7 using the full NSATMOS and 
fixed-gravity NSATMOS($g_s$)
models to lines of constant $\Rinf$ for a blackbody spectrum (or
for any spectrum of the scaled form in equation [B1]).  The contours
derived from our grid of NSATMOS models lie closer to lines of
constant $R_{\infty}$ than do the contours derived from the fixed-gravity 
models,
but there are significant differences.  (See Appendix B for further
discussion of the surface gravity effects.)  

The effect of self-consistently including variation in $g_s$ is quite
substantial, as can be seen from the difference between our NSATMOS
contours and the contours of NSATMOS($g_s$).  
The predictions of the NSA($g_s$) model are very similar to those of
our NSATMOS($g_s$) model, demonstrating that variation in surface
gravity is the primary reason for the differences in the results
obtained with NSATMOS and NSA.  This demonstrates that the NSA($g_s$)
and HYD\_SPECTRA($g_s$) models currently available in XSPEC do not give
 correct constraints in regions of parameter space
where the assumed surface gravities are different than those for which
they were computed, and which are outside their region of validity.  
Explicitly, the constraints on neutron star mass and radius computed by 
HGL03 used these models outside their range of validity, and produced 
incorrect results.  It is possible that this error in assumptions may 
affect the results of \citet{Pons02} and \citet{Walter02} as well.   

Some works have used the NSA models in XSPEC to constrain the range of 
$R_{\infty}$ \citep{Rutledge01a,Rutledge01b,Gendre03a,Gendre03b}.  This 
method is more accurate due to a degeneracy in spectral shape 
variations between surface gravity and surface temperature
 \citep[see ][and Appendix B]{Zavlin98}.  For instance, fitting X7 with the 
NSA model, freezing the mass (1.4 \Msun) and radius (10 km), we 
infer $R_{\infty}=18.34$ km, with a 90\% confidence range between 17.1 and 
22.1 km.  Comparing this to Figure 4, we see that this range is more 
accurate than the method of HGL03, though still inaccurate at the 
$\sim$10\% level.  
The differences between hydrogen atmosphere spectra
and blackbodies are large enough that direct calculation of a range of
$M_{\rm ns}$ and $R_{\rm ns}$ values giving acceptable fits is
preferable to simply constraining $R_{\infty}$.
  Measuring $R_{\infty}$ is generally satisfactory, however, when the observational 
data is not of the highest quality, and/or the purpose of the measurement is 
simply to check for consistency with the canonical NS mass and radius (a  
typical test of current observational data).
 
Applicability of any pure hydrogen-atmosphere model to a NS
atmosphere requires that accretion is not continuing above the critical
rate to keep metals present in the atmosphere, as that would alter the
opacity of the atmosphere \citep{Brown98}.  The X-ray luminosity
produced by the critical accretion rate is similar to the observed
X-ray luminosity of X7.  Thus, it remains conceivable that the
inferred mass and radius of X7 are biased by the presence of metals,
and thus extra opacity, in the NS atmosphere.  However, the
extraordinary stability of X7's X-ray flux, and its lack of accretion
signatures such as a hard power-law spectral component, or features due
to lines or edges, indicate that X7's X-ray emission is most likely
produced by deep crustal heating rather than continued accretion.

\section{Conclusions}

We have computed new grids of hydrogen atmosphere models for neutron stars in
quiescent LMXBs, accounting self-consistently for electron thermal
conduction, radiation force, self-irradiation of the NS, and
variations in the NS surface 
gravity.  The first three effects do not produce large changes, and our 
NSATMOS models agree very well with the NSA models of \citep{Zavlin96} in 
the appropriate surface gravity regime.  We find that the effects
of freely varying the NS surface gravity, on the other hand,
 are significant, especially when
trying to constrain the mass and radius of neutron stars.

We report new (2002) \Chandra observations of the quiescent LMXB known as X7
in 47 Tuc.  No variability is observed on any time scale between
minutes and weeks in the new data set.  The 2002 \Chandra data, plus 
the 2000 \Chandra data, can be simultaneously fit with our hydrogen
atmosphere model, photoelectric absorption, and a correction model for
instrumental pileup.
No convincing spectral features are seen, and no additional hard
component is detected.  No variations are seen in X7's spectrum over the
2.5 year interval.  

Our spectral fitting constrains the range of mass and radius which can
produce a spectrum like that observed.   In contrast to the results
of \citet{Heinke03a}, our use of a range of surface gravities allows
acceptable fits to X7 with a standard 1.4 \Msun NS mass and a radius
of 14.5$^{+1.8}_{-1.6}$ km, as well as a high-mass solution
(M=2.20$^{+0.03}_{-0.16}$ \Msun) for a radius of 10 km.  
For a canonical mass of 1.4
\Msun\, a 10-12 km radius is ruled out at 99\% confidence.  
This indicates (assuming the validity of a pure hydrogen atmosphere model) 
that either X7 is more massive than any NS yet known, or
that X7 has a somewhat larger radius than canonical modern NS
models (our preferred interpretation). In either case a relatively ``stiff'' 
NS equation of state is favored.

The HYD\_SPECTRA \citep{Gansicke02} and NSA hydrogen-atmosphere NS
models as currently implemented in XSPEC do not include a range of
surface gravities appropriate for the possible ranges of NS masses and
radii.  Our work using NSATMOS has shown that accurate constraints 
(at the $<$10\% level) on 
the radius and mass of NSs require self-consistent modeling of the effects of 
surface gravity.  Our NSATMOS model, and the NSAGRAV code 
\citep[using the models of][provided for XSPEC 
during this paper's refereeing process]{Zavlin96}, meet these requirements.

\acknowledgements

We warmly acknowledge Don Lloyd for providing comparison neutron star
   atmosphere models for use in verifying our code, and 
Marc Freitag for the use of his fig2curve script.  We also thank Lars
   Bildsten, Bob Rutledge and Jim Lattimer for useful discussions and
   suggestions, and the CXC 
 team for their untiring data calibration efforts.  This work was supported in
part by NSF grant AST 0307433, and in part by the Lindheimer
   Postdoctoral Fellowship at Northwestern University.

\appendix
\section{The NSATMOS code} 

The NSATMOS code calculates an atmospheric model under the following
conditions and assumptions:
(1) Static, atmosphere in the plane-parallel approximation;
(2) Negligible magnetic fields ($B \lesssim 10^8$ G);
(3) Ideal equation of state for pure hydrogen with complete ionization;
(4) Opacity due to thermal free-free absorption plus Thomson scattering in
the unpolarized, isotropic approximation; 
(5) Energy transport by radiation and electron heat conduction;
(6) Hydrostatic equilibrium includes radiation force due to absorption and
scattering; 
(7) Comptonization is ignored; and
(8) For a compact object within its own photon sphere,$R < (3/2)\Rs$, 
self-irradiation of the surface is taken into account ($\Rs=2GM/c^2$ is
the Schwarzschild radius).
We note that the neglect of neutral
hydrogen limits the validity of the code to temperatures $\Teff
\gtrsim 3 \times 10^5$ K, while the omission of Comptonization
limits it to perhaps $\Teff \lesssim 3 \times 10^6$ K \citep{Zavlin96}.  These
limitations are not serious for the present application, but work
is in progress to correct NSATMOS for both.  

The solution of the atmosphere problem based on these assumptions
follows mostly along standard lines for stellar atmospheres.  Here we
review the main ideas for reference.  

The depth variable used here is the mass column density $m$,
measured from the surface, which is related to vertical height $z$ by the
differential relation $dm=-\rho\,dz$, where $\rho$ is the mass density.  

The intensity field $I_\nu(m,\mu)$ is a function of frequency $\nu$,
depth, and $\mu$, the cosine of the angle relative to the outward
normal.The transfer equation for the intensity is
then 
\be
 \mu \pder{I_\nu}{m} = \kappa_\nu( I_\nu - B_\nu)
            +\kappaT ( I_\nu - J_\nu ).   \eqlab{A2}
\ee
The Planck function is denoted by $B=B_\nu(T)$, where $T=T(m)$ is the
local temperature.  For a fully ionized hydrogen plasma the Thomson
opacity $\kappaT$ is a constant, but the free-free opacity
$\kappa_\nu$ depends on the density and temperature.  The mean
intensity $J_\nu$ is defined by the integral,
\be
       {1 \over 2}\int_{-1}^{+1} \mu^{j} I_\nu \, d\mu,  \eqlab{A1}
\ee
for $j=0$.  Other useful moments $H_\nu$ and $K_\nu$ are defined by this
integral for $j=1$ and $j=2$, respectively.  

When the stellar surface is outside the photon sphere, $R> (3/2)\Rs$,
the surface boundary condition on the intensity field $I_\nu(m,\mu)$
is the usual one of no incident radiation.  When $R < (3/2)\Rs$, the
self-irradiation of the surface due to gravitational bending of the
rays is taken into account as discussed in McClintock et al.\ (2004).
The boundary condition at suffiently large depth is that the radiation
field is given by the LTE diffusion approximation based on the local
gradient of the temperature field.

An important quantity is the radiative monochromatic energy flux,
given in terms of the $H_\nu$ moment by $F_\nu=4\pi H_\nu$.  The total
radiative energy flux is found by integration over frequency,
\be
   F\rad = \int_0^{\infty} F_\nu\, d\nu 
         = 4\pi \int_0^{\infty} H_\nu\, d\nu.  \eqlab{A3}
\ee
For some ranges of parameters of interest, the heat conduction by
electrons can be of importance.  For the conductive flux we use the
Spitzer-H\"arm formula
\be
  F\con = - \Spitzer T^{5/2} \,\pder{T}{z},  \eqlab{A4}
\ee
The constant $\Spitzer$ has a standard value of order $1.8 \times
10^{-5}$ erg cm$^{-1}$ s$^{-1}$ K$^{-7/2}$.
 Magnetic fields tend to suppress the thermal conduction, though by only 
a factor of order 3-5 when the field is chaotically tangled  
 \citep{Narayan01}.  Therefore, for the calculations of this paper we have 
assumed a conductivity equal to a third of the Spitzer value.  However, we 
find that electron conduction even at the Spitzer value is a
negligible contributor to the energy flux, making the issue moot.
We take as a boundary condition at the surface that the conductive flux
vanishes, $F\con(0)=0$, since no electrons exist above the outer
surface.

Energy equilibrium requires that the net outward energy flux, radiative plus
conductive, be constant with depth,
\be
    F\tot= F\rad +F\con  = \sigma \Teff^4.  \eqlab{A5}
\ee
The constant total flux is parametrized by the {\em effective
temperature} $\Teff$.  Since the conductive flux vanishes at the
surface, the effective temperature is, as usual,a measure of the
total emitted radiative flux.  

Equation (\eqref{A5}) is used in NSATMOS as the basic expression of
energy equilibrium.  We note that another common method of expressing
this is through the flux derivative condition,
\be
    {dF\tot \over dm} = 0.  \eqlab{A5'}
\ee
If this form is used, the desired value of total flux $F\tot$ must be
introduced in some other way, usually through the boundary conditions
at depth on radiation and conduction.

The hydrostatic equilibrium equation is
\be
  \pder{p}{m} = g - \grad, \eqlab{A6}
\ee
where $p$ is the gas pressure and $g$ is the local acceleration of
gravity, given by the usual formulas in Schwarzschild geometry.  The
radiative weakening of gravity $\grad$ is given by the radiation force
per unit mass due to free-free absorption and Thomson scattering,
\be
     \grad = {1 \over c}\int (\kappa_\nu+\kappaT)F_\nu\, d\nu. \eqlab{A7}
\ee

In a completely ionized hydrogen plasma, the electron and proton densities
are equal, $\Ne=\Np$, and the ideal equation of state is
\be
    p=2\Ne kT   \eqlab{A8}
\ee
where $k$ is the Boltzmann constant.

The model atmosphere problem requires the simultaneous,
self-consistent solution of the preceding equations for the radiation
field and the gas properties (temperature, density, and pressure) as
functions of depth.  Overall, this is a set of nonlinear, coupled
equations.  Our method of solution involves (as do all such methods) a
preliminary recasting of the equations into favorable forms such that
cycles of iteration between them converge reasonably rapidly to the
desired solution.

One very useful trick is to recast the transfer equation in terms of
Eddington factors.  Multiplication of equation (\eqref{A2}) by 1 and
$\mu$, followed by integration over all $\mu$ gives
\bea
    &&\pder{H_\nu}{\tau_\nu} =\epsilon_\nu (J_\nu - B_\nu), \eqlab{A9}\\
    &&\pder{K_\nu}{\tau_\nu} = H_\nu,    \eqlab{A10}
\eea
where the monochromatic optical depth is defined differentially as
$d\tau_\nu = (\kappa_\nu+\kappaT)\, dm$ with $\tau_\nu=0$ at the surface,
and where
\be
    \epsilon = {\kappa_\nu \over \kappa_\nu+\kappaT}. \eqlab{A11}
\ee

Introducing the Eddington factor $f_\nu=K_\nu/ J_\nu$,
equations (\eqref{A9}) and (\eqref{A10}) can be expressed as the
single second-order equation,
\be
     {\partial^2 (f_\nu J_\nu) \over \partial \tau_\nu^2} 
          =\epsilon_\nu (J_\nu - B_\nu), \eqlab{A12}
\ee
The two boundary conditions on this equation at the surface and at
depth are expressed through the boundary Eddington factors $g_{\nu 0}$
and $g_{\nu d}$, such that,
\be
     H_\nu(0)=g_{\nu 0}J_\nu(0), \qquad\qquad  
     H_\nu(\tau_{\nu d})=g_{\nu d}J_\nu(\tau_{\nu d}).   \eqlab{A13}
\ee
After a set of initial Eddington factors is chosen, they are updated
during the course of iterative solution using full angular formal
solutions of the transfer equation (\eqref{A2}).  It is during this
formal solution that the detailed boundary conditions on intensity
come into play, including the possibility of sef-irradiation.

A particularly critical part of the iterative solution is how the run
of temperature with depth $T(m)$ is corrected after each iteration.
In doing this it is important to take account of how radiation can
couple the variables at distant points in the atmosphere.  Methods
that try to correct the temperature based on local information, or
even on information just a few optical depths away, will converge much
too slowly to be practical.

Here we adopt a temperature correction scheme based on a partial
linearization of the temperature dependence in the transfer equation
and the energy equation.  The temperature field is
represented as
\be
     T(m) = T\zth(m)+T\fst(m),   \eqlab{A14}
\ee
where $T\zth(m)$ is an initial ``guess'' for the temperature, and $T\fst(m)$
is the ``correction.''  In the perturbation sense these are regarded as
of zeroth and first order, respectively.  Other quantities in the
problem can be expanded to first order in the perturbation.  
For example, the radiation field has the representation,
\be
    I_\nu=I_\nu\zth+I_\nu\fst.  \eqlab{A15}
\ee
Not all quantities in the problem are expanded in this way, making
this a partial linearization rather than a complete linearization
scheme.  The decision as to which quantities are expanded and which
are not is somewhat arbitrary, but is guided by the physical idea that
the critical equations are the energy equation, the transfer equation,
and the conduction equation, since these are the ones that control the
global redistribution of energy in the atmosphere.

The zeroth order form of the transfer equation (\eqref{A12}) is
\be
     {\partial^2 (f_\nu J_\nu\zth) \over \partial \tau_\nu^2} 
          =\epsilon_\nu [J_\nu\zth - B_\nu(T\zth)]. \eqlab{A16}
\ee
We take as the first order form,

\be     {\partial^2 (f_\nu J_\nu\fst) \over \partial \tau_\nu^2} 
          =\epsilon_\nu [J_\nu\fst - \Bdot_\nu(T\zth)T\fst], \eqlab{A17}
\ee
where $\Bdot(T) = \partial B_\nu(T)/\partial T$.  Note that the
radiation fields and the temperature dependence of the Planck function
have been expanded, but not the Eddington factors or the opacities, so
that the optical depth scale is unchanged.

Expansion of the energy equation (\eqref{A5}) through first order
gives, after some rearrangement,
\be
  F\tot\fst=F\rad\fst + F\con\fst =
   \sigma \Teff^4 - ( F\rad\zth + F\con\zth)=-E.  \eqlab{A18}
\ee
 The right hand side is the negative of the flux ``error'' $E$ in the
zeroth order solution.  The left hand side is the total first order flux,
consisting of the two terms,
\bea
   F\rad\fst &=& 4\pi \int_0^{\infty} H_\nu\fst \, d\nu,  \eqlab{A19}\\
   F\con\fst &=& - \Spitzer \pder{}{z}\left[(T\zth)^{5/2}T\fst \right]
          \eqlab{A20}
\eea
Each of these depends linearly on the first order temperature
correction $T\fst$.  For $F\con\fst$ this follows directly from equation
(\eqref{A20}).  For $F\rad\fst$ we note that  $H\fst$
depends linearly on $J_\nu\fst$, since
\be
     \pder{(f_\nu J_\nu\fst)}{\tau_\nu} = H_\nu\fst,  \eqlab{A21}
\ee
which follows from equation (\eqref{A10}), noting $K_\nu\fst=f_\nu
J_\nu\fst$.  Finally, $J_\nu\fst$ depends linearly on $T\fst$ through
equation (\eqref{A17}.

In terms of $N_D$ discretized depths, the linear dependence of
$F\rad\fst(m)$ and $F\con\fst(m)$ on $T\fst(m)$, can be expressed as
\bea
  F\rad\fst &=& \Phi\rad \cdot T\fst, \eqlab{A22}\\
    F\con\fst &=& \Phi\con \cdot T\fst,  \eqlab{A23}
\eea
where $\Phi\rad$ and $\Phi\con$ are finite matrix operators over the
discrete depths.

In order to construct the matrix $\Phi\rad$, $N_F$ discrete frequencies are
introduced.  For each of these frequencies,
the transfer equation (\eqref{A16}), plus its boundary conditions,
becomes a tridiagonal matrix equation in depth, which may be solved in
of order $N_D^2$ operations.  In this way one obtains matrices $A_\nu$
such that $J_\nu\fst= A_\nu \cdot T\fst$ for each frequency.
Implementing equation (\eqref{A21}) with some form of numerical
differentiation, we obtain matrices $B_\nu$ such that $F_\nu\fst=
B_\nu \cdot T\fst$.  Implementing a frequency quadrature scheme for
equation (\eqref{A19}), we then obtain the matrix $\Phi\rad$.  The
whole process takes of order $N_F N_D^2$ operations.

The construction of the matrix $\Phi\con$ is easier, since it merely
involves implementing a numerical differentiation formula for equation
(\eqref{A20}) plus boundary condition.  Depending on the order of the
formula, the resulting matrix might be bidiagonal or tridiagonal.

Defining $\Phi\tot = \Phi\rad + \Phi\con$, equations (\eqref{A18}),
(\eqref{A22}), and (\eqref{A23}) can be combined to give,
\be
    \Phi\tot \cdot T\fst =-E,   \eqlab{A24}
\ee
which relates the temperature correction $T\fst$ directly to the
flux error $E$ in the trial (zeroth order) solution.  This matrix equation
is then solved for $T\fst$, and a new trial temperature law is obtained
from $T\zth \leftarrow T\zth + T\fst$.  

Using the new trial temperature, improved radiation fields, Eddington
factors, radiative force, and hydrostatic equilibrium are computed.
This process is repeated until convergence is obtained.

This method was used to compute our grid of models.  In order to
initialize the first model, we used a Rosseland-type solution for the
run of temperature (similar to the asymptotic result given in equation
(15) of Zavlin et al.\ 1996) and we set the radiative force to zero.
For subsequent models we adopted those quantities from a previously
calculated model with nearby parameters.  This was particularly
helpful in computing models close to the Eddington limit.

Some additional ``tricks'' we found useful for some models were: (1)
Adiabatic turning on of parameters, in which the model parameters were
changed in small steps from those of a previously solved case to the
desired ones, these changes occuring along with the other iterations;
and (2) Linearization limiters, which are virtually identical to
(\eqref{A14}) for small corrections, but which keep the temperature
correction within fixed bounds, for example,
\be
      T(m) = T\zth(m) + {T\zth(m) \over 2} \tanh 
               \left[ {2 T\fst(m) \over T\zth(m)} \right]. \eqlab{A25}
\ee

\section{Surface Gravity Effects}  

Our spectral fitting to constrain X7 in the mass-radius plane has
shown the importance of considering neutron star models with a full
range of surface gravities as well as effective temperatures.  This
conclusion is based on numerical results of the XSPEC analysis
program, which involves a long chain of data reductions plus
manipulations of the the model atmosphere spectra.  In this appendix we
attempt to understand better the importance of the surface gravity
$\gs$ using some heuristic, qualitative arguments.

We begin by considering a approximate analytical representation for
the surface spectral flux of the NS atmosphere, namely,
\be
     F_\nu(\nu)=\Teff^3 \phi(\xi),\qquad\qquad \xi=\nu/\Teff,   \eqlab{R1}
\ee
where $\phi(\xi)$ is a fixed function.  This implies that there is a single
spectral shape which is simply rescaled by the effective temperature $\Teff$.
Such a representation would hold exactly if the neutron star surface
radiated as a blackbody.  It also holds approximately for more
realistic pure ionized hydrogen atmospheres; an approximate expression
of this form was given in equation (A17) of McClintock et al.\ (2004).

Given the spectral flux at the surface, the observed spectral flux
$\Fobs(\nuobs)$ at a large distance $D$ may be expressed (see,
e.g., McClintock et al.\ 2004).  Let us assume for the moment that the
surface flux is given by (\eqref{R1}) exactly.  Then for a neutron
star with mass $\Mns$ and radius $\Rns$ we may write,
\be
   \Fobs(\nuobs) =  {\Rinf^2 \over D^2} \Tinf^3 \phi (\nuobs/\Tinf), 
         \eqlab{R2}
\ee
where $\Tinf=\Teff/(1+z)$, $\Rinf=\Rns(1+z)$.  Here $1+z =
(1-\RS/\Rns)^{-1/2}$ is the gravitational redshift factor at the
neutron star surface, and $\RS=2G\Mns/c^2$ is the Schwarzschild radius.

\begin{figure}[t] 
\epsscale{1.0}
\plotone{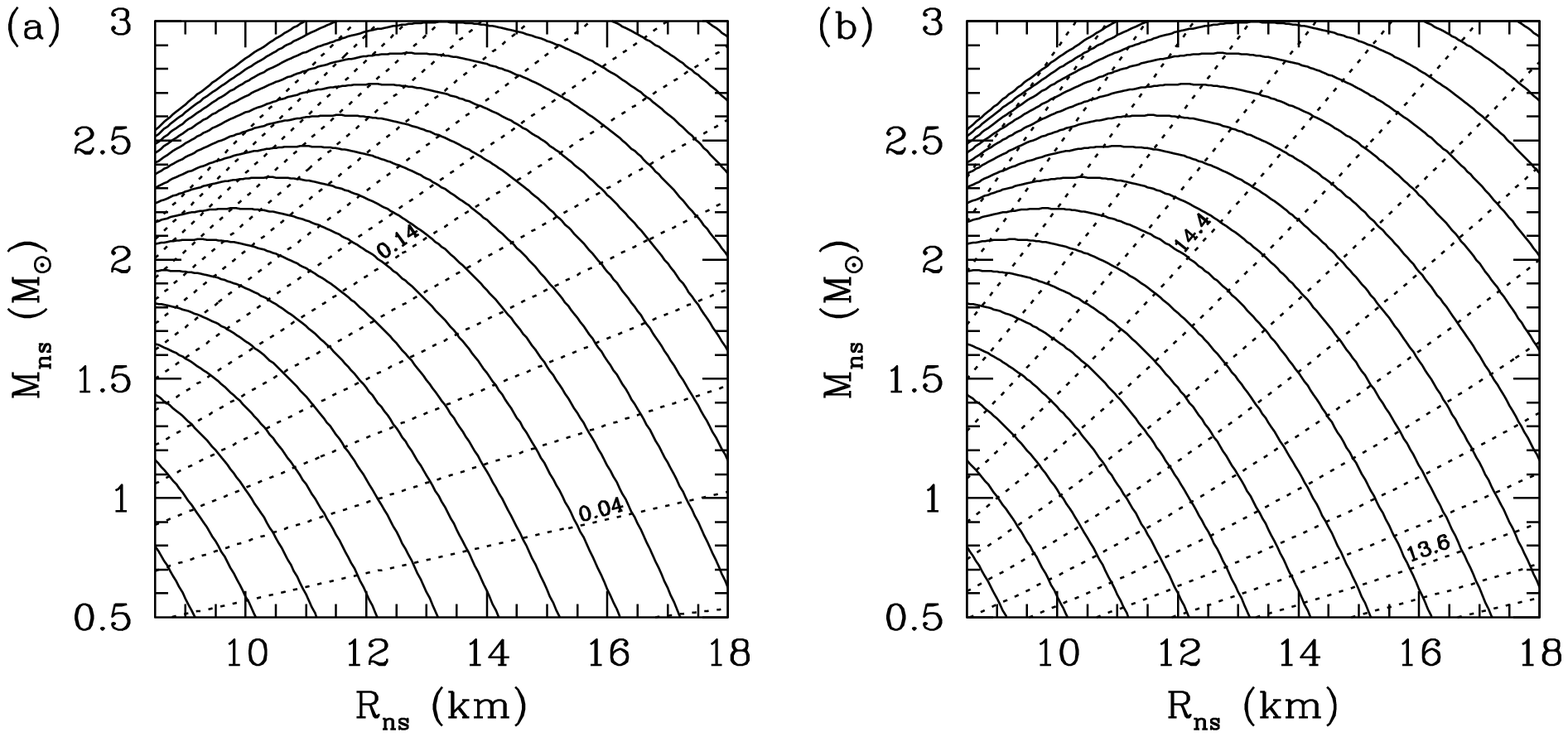}
\caption{(a) The $\Mns$--$\Rns$ plane, showing solid curves of constant 
  $\Rinf$ and dashed curves of constant $(1+z)$. Two of these 
  dashed curves are labelled for the values $\log (1+z)=0.04$ and 
  $0.14$.  The others
  are at logarithmic spaced values with $\Delta \log (1+z)=0.02$.
  (b) The same, except with dashed curves of constant surface gravity $\gs$.
  The two marked curves are for $\log\gs = 13.6$ and $14.4$, with other
  curves spaced with $\Delta\log\gs = 0.1$.
     \label{fig:Rfig05}}  
\end{figure}

With a prescribed form of the function $\phi$ and a known distance
$D$, spectral fitting is reduced to the determination of just two
parameters, $\Tinf$ and $\Rinf$.  However, there are three parameters
required to specify a neutron star model fully, which can be taken to
be its mass $\Mns$, radius $\Rns$, and effective temperature $T$.
This implies that for each fit there is a one-parameter set of models
that are observationally indistinguishable from that fit.  
To show this in detail, we first note that the relation
$\Rinf=\Rns(1+z)$ can be expressed,
\be
  \Mns = \left(1-{\Rns^2 \over \Rinf^2} \right) {c^2 \Rns \over 2G}.  \eqlab{R3}
\ee
An acceptable $\Mns$--$\Rns$ pair can lie anywhere on one of the
contour lines of constant $\Rinf$, shown as the solid curves in figure
\ref{fig:Rfig05}(a).  
%
%
\placefigure{fig:Rfig05}
The additional parameter along each curve can
be taken to be the ratio $\Teff/\Tinf$, which is equal to the redshift
factor $(1+z)$, given in terms of $\Rns$ and $\Mns$ by
\be
     1+z = \left( 1- {2G\Mns \over c^2 \Rns} \right)^{-1/2}.  \eqlab{R4}
\ee
Curves of constant $(1+z)=\Teff/\Tinf$ are plotted as dashed curves in
figure \ref{fig:Rfig05}(a).  However, since there is no independent
determination of $\Teff$, this relation does not constrain the possible
locations along the contours of $\Rinf$.  In a spectral fitting program
such as XSPEC, this degeneracy would show itself by the fact that the
confidence contours would be made up of contours of constant $\Rinf$,
that is, the solid curves in figure \ref{fig:Rfig05}(a).

In practice, since the simple scaling law (\eqref{R1}) is only
approximately valid, the confidence contours will not follow those
solid curves precisely.  Nonetheless, one can see in figures \ref{fig:rybcont},
\ref{fig:allcont}, and \ref{fig:rinf}
that the confidence contours are generally consistent with there being
a blackbody-like degeneracy along their length, since they do not close
off on either ends of the confidence bands.  One also sees that the
shapes of the confidence contours are different depending on
whether the gravity has been fixed or allowed to vary freely.  The
fixed gravity curves can be characterized as being pushed out to
larger radii at smaller mass, giving the curves in this region a
flatter appearance.  The curves for freely varying gravity do the
opposite: they are pushed to smaller radii at smaller mass, so much so
that they actually curl around and change their direction. (It is this
fact that allows a fit with a smaller radius, one of the major results
of this paper.)

We can give a heuristic, qualitative argument for this difference.  
First, let us note that the surface gravity, which is required for the
construction of the atmosphere, is given by,
\be
       \gs={G\Mns (1+z) \over \Rns^2}.   \eqlab{R5}
\ee
In figure \ref{fig:Rfig05}(b), curves of constant $\Rinf$ (solid) are again
plotted in the $\Mns$--$\Rns$ plane, but now with curves of constant 
$\gs$ (dashed).

\begin{figure}[t] 
\epsscale{1.0}
\plotone{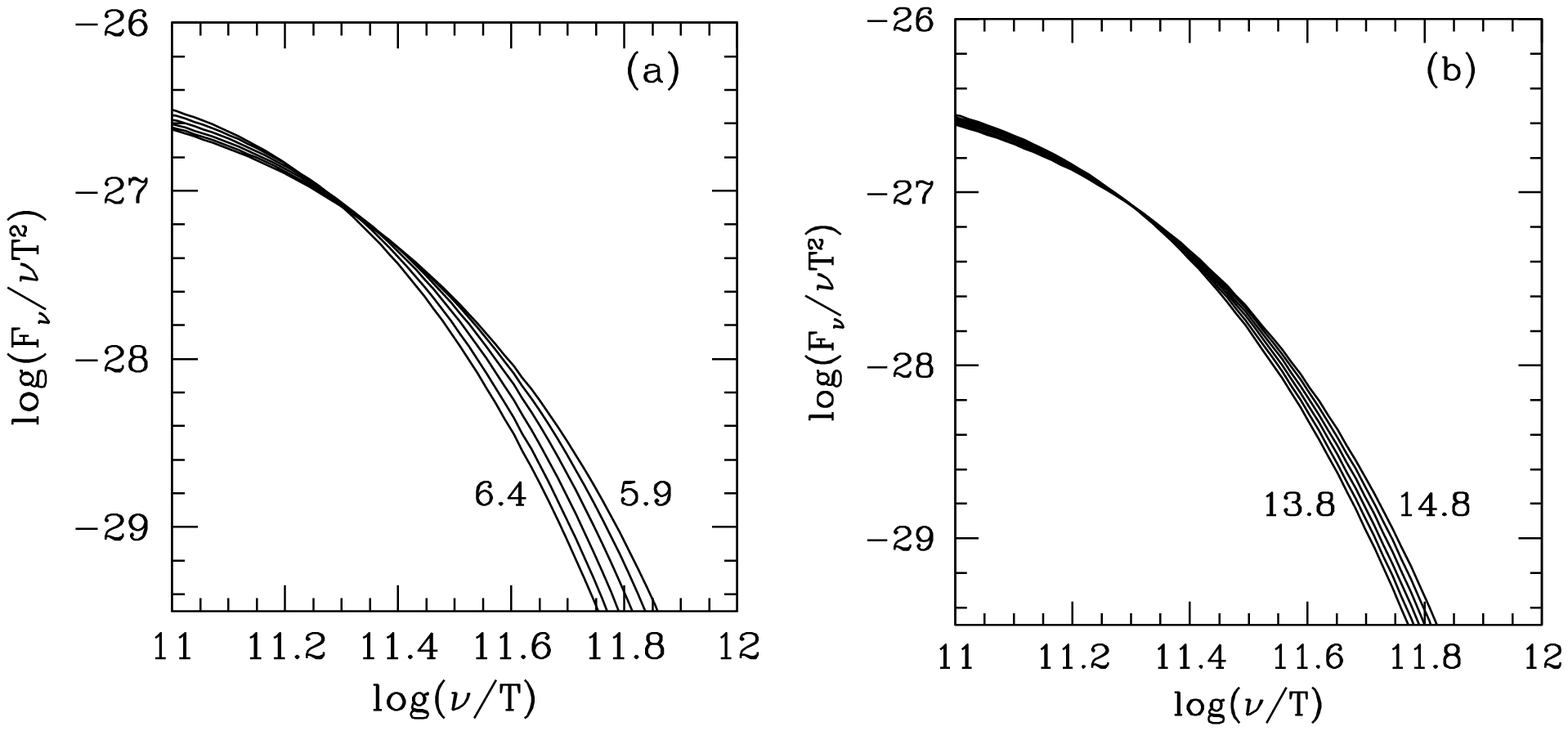}
\caption{(a) Scaled surface number flux vs.\ scaled frequency for 
   $\log\gs = 14.4$ and for $\log\Teff$ ranging from 5.9 to 6.4.  
(b) Scaled surface number flux vs.\ scaled frequency for $\log \Teff = 6.2$
and for $\log\gs$ ranging from 13.8 to 14.8.
     \label{fig:Rfig06}}  
\end{figure}

The sensitivity of the function $\phi(\nu/\Teff)$
to temperature and gravity is demonstrated in figure \ref{fig:Rfig06}.
%
%
\placefigure{fig:Rfig06}
Here is plotted the quantity $\log (F_\nu/\nu\Teff^2)$ versus
$\log(\nu/\Teff)$ for the surface fluxes of the neutron star models
for various effective temperatures [\ref{fig:Rfig06}(a)] and various
gravities [\ref{fig:Rfig06}(b)] in the neighborhood of the approximate
fitted values $\log\Teff=6.2$ and $\log\gs = 14.4$.  The frequency
range here was chosen to match roughly the portion of the flux curves
observed in X7.  If the scaling relation (\eqref{R1}) were exact,
these would all lie on the same curve, namely, $\log[\phi(\xi)/\xi]$.
In fact, there is a sensitivity to both effective temperature and
gravity.  
We see that the flux distribution steepens for increasing temperature, 
and for decreasing surface gravity.  
The sensitivity to temperature is greater than that of gravity, in the
sense that 0.5 dex variation of temperature gives a substantially
greater variation than 1.0 dex variation in gravity.  However, an
essential point to notice from figures \ref{fig:Rfig05}(a) and
\ref{fig:Rfig05}(b) is that over the part of the $\Mns$--$\Rns$ plane
where the spectral fitting results are most divergent, say, for $0.5 <
\Mns/\Msol < 2 $ and for $12
\hbox{ km} < \Rns < 18$ km, the gravity changes by about 1.0 dex,
while the temperature changes by about only 0.1 dex [a range
corresponding to two neighboring curves in figure \ref{fig:Rfig06}(a)].
In that case, the change due to gravity is effectively more
substantial than that due to temperature.  Also, the changes in the
curves are of the opposite sense as one moves along the curves of
constant $\Rinf$ from the top to bottom of the plot, since both
temperature and gravity decrease.  Thus, if one fixes the gravity, the
confidence contours are pushed in one way by the sensitivity to
temperature, but if gravity is allowed to vary, then the larger
effective sensitivity to gravity pushes the confidence curves in the
opposite direction.  The preceding heuristic argument accounts
qualitatively for the overall behavior seen in figures \ref{fig:rybcont},
\ref{fig:allcont}, and \ref{fig:rinf} and
demonstates the need to include variable gravity when doing spectral
fitting for such objects.

\bibliography{src_ref_list}
\bibliographystyle{apj}

\clearpage




%
%
%


%
%
%


%

%
%

\clearpage  

%

%
%
\clearpage  

\clearpage  

\end{document}